\documentclass[aps,preprint,amssymb,12pt,floatfix]{revtex4}
\usepackage{subfigure}
\usepackage{graphicx}
\usepackage{rotating}
\usepackage{tabularx}
\usepackage{mciteplus}

\def\la{\langle}
\def\ra{\rangle}

\begin{document}

\title{Molecular origin of constant $m$-values, denatured state 
collapse, and residue-dependent transition midpoints in globular proteins}

\renewcommand{\thefootnote}{\fnsymbol{footnote}}
\author{Edward P. O'Brien$^{1,2}$, Bernard R. Brooks$^{2}$, and D. Thirumalai$^{1}$\footnote{Corresponding author:
Institute for Physical Science and Technology, University of Maryland,
College Park, MD 20742, phone: 301-405-4803; fax: 301-314-9404;
e-mail: thirum@umd.edu}}
\affiliation{\small $^{1}$Biophysics Program, Institute for Physical
Science and Technology\\
and Department of Chemistry and Biochemistry\\
University of Maryland, College Park, MD 20742\\
$^{2}$Laboratory of Computational Biology\\
National Heart Lung and Blood Institute\\
National Institutes of Health, Bethesda, MD 20892}
\vspace{2cm}

\date{\small \today}
\maketitle
\baselineskip = 20pt
\newpage

\noindent
\textbf{Title Running Head:} Origin of protein $m$-values, collapse, and individual residue behavior
\\
\\
\textbf{Abbreviations:} TM, Transfer Model; MTM, Molecular Transfer Model; D, Denatured state; N, Native state; DSE, 
Denatured state ensemble; NSE, Native state ensemble; $C_{\alpha}$-SCM, $C_{\alpha}$ side chain model; MREX, Multiplexed
Replica Exchange; GdmCl, Guanidinium Chloride; NMR, Nuclear Magnetic Resonance; FRET, Forster Resonance Energy Transfer.
\\
\\
\textbf{Funding:}
This work was supported in part by a grant from the NSF (05-14056)
and Air force office of scientific research (FA9550-07-1-0098)
to D.T., a NIH GPP Biophysics Fellowship to E.O., and by the Intramural
Research Program of the NIH, National Heart Lung and Blood Institute.

\newpage
\textbf{Abstract}

Experiments show that for many two state folders the
free energy of the native state $\Delta G_{ND}([C])$ changes linearly as the denaturant concentration
[C] is varied. The slope, $m= \frac{d\Delta G_{ND}([C])}{d[C]}$, is nearly constant.
According to the Transfer Model,
the $m$-value is associated with the difference in the surface area between the
native (N) and the denatured (D) state, which should be a function of $\Delta R_g^2$, the difference in the 
square of the radius of gyration between
the D and N states. Single molecule experiments show that $R_g$ of the structurally heterogeneous denatured
state undergoes an equilibrium collapse transition as [C] decreases, which
implies $m$ also should be [C]-dependent. We resolve the conundrum
between constant $m$-values and [C]-dependent changes in $R_g$ using molecular
simulations of a coarse-grained representation of protein L, and the Molecular Transfer
Model, for which the equilibrium folding can be accurately calculated as a function
of denaturant (urea) concentration.
In agreement with experiment, we find that over a large range of denaturant
concentration ($>3$ M) the $m$-value is a constant, whereas under strongly
renaturing conditions ($<3$ M) it depends on [C].
The $m$-value is a constant above [C]$>3$ M because the [C]-dependent changes in the surface area of the backbone groups, which  make the largest
contribution to $m$,  is relatively narrow in the denatured state. The burial of the
backbone and hydrophobic side chains gives rise to substantial surface area
changes below [C]$<3$ M, leading to collapse in the denatured state of protein L.
Dissection of the contribution of various amino acids to the total surface area
change with [C] shows that both the sequence context and residual structure are important. There are  [C]-dependent variations
in the surface area for {\it{chemically identical}} groups such as the backbone or Ala.
Consequently, the midpoint of transition of individual residues vary significantly
(which we call the Holtzer Effect) even though global folding can be described
as an all-or-none transition.
The collapse is specific in nature, resulting in the formation
of  compact structures with appreciable populations of native-like secondary structural
elements.
The collapse transition is driven by the loss
of favorable residue-solvent interactions and a concomitant increase in the strength of
intrapeptide interactions with decreasing [C]. The strength
of these interactions is non-uniformly distributed throughout the
structure of protein L. Certain secondary structure elements have
stronger [C]-dependent interactions than others in the denatured state.
\newpage


\noindent
The folding of many small globular proteins is often modeled using the two-state approximation 
in which a protein is assumed to exist in either the native (N) or the denatured (D) states \cite{JacksonFD1998}. 
The stability of N relative to D, $\Delta G_{ND}(0)$, is typically obtained by measuring $\Delta G_{ND}([C])$
as a function of the denaturant concentration [C], and extrapolating to 
[C]=0 using the linear extrapolation method (LEM) \cite{SantoroBIOCHEM1992}. The denaturant-dependent 
change in native state stability, $\Delta G_{ND}([C])$, for these globular proteins is usually a 
linear function of [C] \cite{GreeneJBC1974,PaceMenz1986,BolenBIOCHEM1988,FershtBIOCHEM1991,
SantoroBIOCHEM1992,MakhatadzeJPC1999,FershtBook,RosePNAS2006}. 
Thus, $\Delta G_{ND}([C]) = \Delta G_{ND}(0) + m[C]$,
where $m = \partial\Delta G_{ND}([C])/\partial[C]$ is a constant \cite{BolenBIOCHEM1988},
which by convention is referred to as the $m$-value.  
However, deviations from linearity, 
especially at low $[C]$, have also been found \cite{BakerFD1997}, indicating that 
the $m$-value is concentration dependent. In this paper we address two inter-related
questions: (1) Why are $m$-values constant for some proteins, even though
there is a broad distribution of conformations in the denatured state ensemble (DSE)? (2) What
is the origin of denatured state collapse, that is the compaction of the DSE, with decreasing
[C] that is often associated with non-constant
$m$-values \cite{RoderBIOCHEM1993,BakerFD1997,BakerBIOCHEM1997}?

\begin{figure}[ht]
\subfigure[]{
  \label{}
  \includegraphics[width=4.5in]{fig1a.eps}
} \\
\vspace{1cm}
\subfigure[]{
  \label{}
  \includegraphics[width=4.5in]{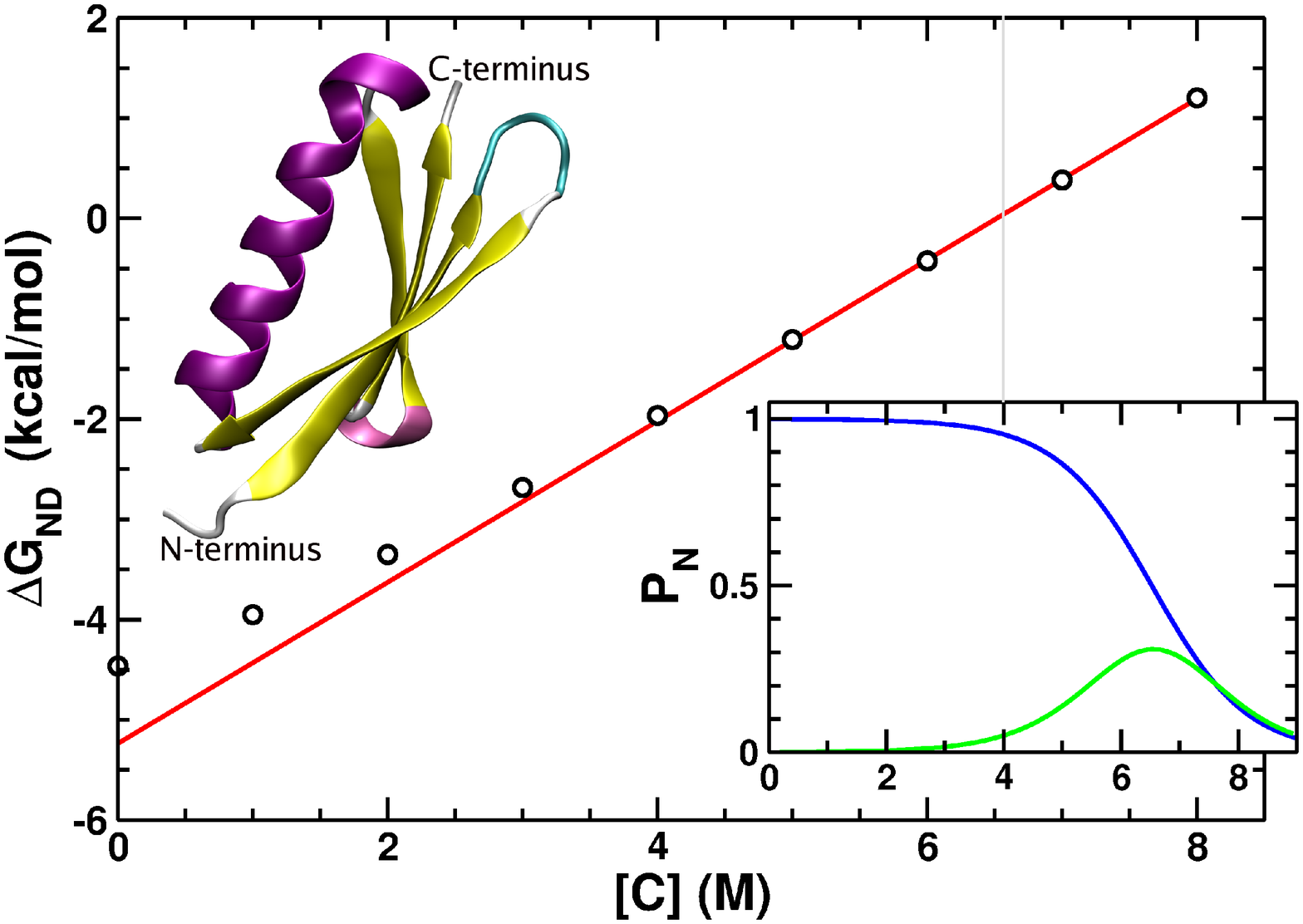}
}\\
\caption{}
\label{dGns}
\end{figure}

Potential answers to the first question can be gleaned by considering the empirical
Transfer model (TM) \cite{TanfordJBC1963,TanfordJACS1964,BolenPNAS2007}, which 
has been remarkably successful in accurately predicting $m$-values for a large number 
of proteins \cite{AutonPNAS2005,BolenPNAS2007}.  The revival in the TM as a practical
tool in analyzing the effect of denaturants (and more generally osmolytes) comes from
a series of pioneering studies by Bolen and coworkers \cite{BolenBIOCHEM2004,AutonPNAS2005,BolenPNAS2007}.
Assuming that proteins exist in only two states \cite{BolenPNAS2007,FershtBook}, 
the TM expression for the $m$-value is
\begin{eqnarray}
m &=& \frac{1}{[C]}\sum_{k=1}^{N_{S}} \frac{n_k\delta g_{k}^{S}([C])}{\alpha_{k,G-k-G}^{S}}\Delta \alpha_k^{S} +
\frac{1}{[C]}\sum_{k=1}^{N_{B}} \frac{n_k\delta g_{k}^{B}([C])}{\alpha_{k,G-k-G}^{B}}\Delta \alpha_k^{B}, \label{m1}
\end{eqnarray}
where the sums are over the side chain (S) and backbone (B) groups of the different amino acid types 
(Ala, Val, Gly, etc.), 
$n_k$ is the number of amino acid residues of type $k$ in the protein, and $\delta g_{k}^{S}$ 
and $\delta g_{k}^{B}$ are the experimentally measured transfer free energies for
$k$ \cite{TanfordJBC1963,BolenBIOCHEM2004,BolenME2007} (Fig. \ref{dGns}a).
In Eq. \ref{m1}, $\Delta \alpha_k^{P} = \la\alpha_{k,D}^{P}\ra-\la\alpha_{k,N}^{P}\ra$ ($P= S$ or $B$), where 
$\la \alpha_{k,D}^{P}\ra$ and $\la\alpha_{k,N}^{P}\ra$ are the average solvent accessible surface areas \cite{RichardsJMB1971} of 
group $k$ in the $D$ and $N$ states respectively, and $\alpha_{k,G-k-G}^{P}$ is the 
corresponding value in the 
tripeptide glycine-$k$-glycine. There are two fundamentally questionable assumptions in the TM model: (1) The 
free energy of transferring a protein from water to aqueous denaturant
solution at an arbitrary $[C]$ may be obtained as a sum of transfer energies of individual groups of
the protein without regard to the polymeric nature of proteins.
(2) The surface area changes $\Delta \alpha_k^P$ are independent of [C],
residual denatured state structure,
and the amino-acid sequence context in which $k$ is found. 

The linear variation of $\Delta G_{ND}([C])$ as [C] changes can be rationalized if
(i) $\delta g_k^P([C])$ is directly proportional to [C], and (ii) $\Delta \alpha_k^P$
is [C]-independent. Experiments have shown that $\delta g_k^P([C])$ is a 
linear function of [C] \cite{MakhatadzeJPC1999} while the near-independence of $\Delta \alpha_k^P$ on [C] can
only be inferred based on the accuracy of the TM in predicting the $m$-values \cite{AutonPNAS2005,BolenPNAS2007}.
In an apparent contradiction to such an inference, small angle X-ray scattering 
experiments \cite{ReganPRSCI1996,Doniach1995JMB,Matthews2006COSB,BilselJMB2007} 
and single molecule FRET experiments \cite{HaasBIOCHEM2001,UdgaonkarJMB2005,Nienhaus2005PNAS,Haran2006,eatonPNAS2007,SchulerPNAS2007}
show that the denatured state properties, such as the radius of gyration $R_g$ and the end-to-end distance ($R_{ee}$),
can change dramatically as a function of $[C]$. These observations suggest that 
the total solvent accessible surface area of the protein, $\Delta \alpha_T$($=\sum_{k=1}^{N_{S}}\Delta \alpha_k^{S}  
+ \sum_{k=1}^{N_{B}}\Delta \alpha_k^{B}$), and the various groups
must also be a function of [C], since we expect that 
$\Delta \alpha_T$ must be a monotonically increasing function of $\Delta R_g^2$, which is the difference between $R_g^2$  
of the D and N states \cite{Nienhaus2005PNAS,VendruscoloPS2007}. 
For compact objects $\Delta \alpha_T \propto \Delta R_g^2$ but for fractal structures the
relationship is more complex \cite{PappuBJ2007}. 
Furthermore, 
NMR measurements have found that many proteins adopt partially structured or random coil-like
conformations at high [C] \cite{FesikJMB1994,DobsonSCI2002,NallBIOCHEM1991,AnasariBIOCHEM2003},
which necessarily have large fluctuations in global properties such as $\Delta \alpha_k^P$ and $R_g$.
Thus, the contradiction between the constancy of $m$-values and the sometimes measurable changes in denatured state
properties is a puzzle that requires a molecular explanation.

Bolen and collaborators have already shown that quantitative estimates 
of $m$ can be made by using measured transfer free energies of transfer free energies of individual groups \cite{AutonPNAS2005,BolenPNAS2007}. 
More importantly, these studies established  the dominant contribution to $m$ arises from the backbone \cite{AutonPNAS2005,BolenPNAS2007}. However, only
by characterizing the changes in the distribution of $\Delta \alpha_k^S$ and $\Delta \alpha_k^B$
as a function of [C] can the reasons for success of the TM in obtaining the global property $m$ be fully appreciated.
This is one of the goals of the present study. In addition, we correlate $m$ with denatured 
state collapse, [C]-dependent changes in residual structure, and the solution forces acting on the 
denatured state - properties that cannot be analyzed using the TM.

The denatured, and perhaps even the native state should be described as ensembles of fluctuating conformations,
and will be referred to as the DSE and NSE (native state ensemble),
respectively.  As a result, it is crucial to characterize the distribution of various molecular properties in these ensembles and how they change with [C] in order to describe quantitatively the properties of the DSE. Because the $D$ state is an ensemble
of conformations with a distribution of accessible surface areas, Eq. \ref{m1} should be considered 
an approximate expression for the $m$-value. Even if the basic premise of the TM is valid,
we expect that $\Delta \alpha_k^P$ should depend on the conformation of the protein and the denaturant
concentration. 
Consequently, the $m$-value should be written with an explicit concentration dependence as
\begin{eqnarray}
m([C]) &=& \frac{1}{[C]}\sum_{k=1}^{N_{S}} \frac{n_k\delta g_{k}^{S}([C])}{\alpha_{G-k-G}^{S}}\left\{\la\alpha_{k,D}^{S}([C])\ra - \la\alpha_{k,N}^{S}([C])\ra\right\} \nonumber
\end{eqnarray}
\begin{eqnarray}
&& + \frac{1}{[C]}\sum_{k=1}^{N_{B}} \frac{n_k\delta g_{k}^{B}([C])}{\alpha_{G-k-G}^{B}}\left\{\la\alpha_{k,D}^{B}([C])\ra - \la\alpha_{k,N}^{B}([C])\ra\right\} \label{m2}
\end{eqnarray}
where $\la\alpha_{k,j}^{P}([C])\ra = \int_{0}^{\infty}\alpha_{k,j}^{P}P(\alpha_{k,j}^{P};[C])d\alpha_{k,P}$ ($j =$ $D$ or $N$ and $P =$ $S$ or $B$). 
In principle, the denominator in Eq. \ref{m2} should also be [C]-dependent, however,
we ignore this for simplicity.
In contrast to Eq. \ref{m1}, the conformational
fluctuations in the DSE and NSE are taken into account in Eq. \ref{m2} by integrating over the distribution of surface areas
($P(\alpha_{k,j}^P;[C])$). Moreover, we do not assume that the surface area distributions are
independent of [C] as is done in Eq. \ref{m1}. Such an assumption 
can only be justified by evaluating $P(\alpha_{k,j}^P;[C])$ using molecular simulations or experiments.

We use the Molecular Transfer Model (MTM) \cite{ObrienPNAS2008} in conjunction with
coarse-grained simulations of protein L using the $C_{\alpha}$ side chain model
($C_{\alpha}$-SCM) (see Methods) to test the molecular origin of the constancy
of $m$-values. Because the conformations
and energies are known exactly in the $C_{\alpha}$-SCM simulations, we
can determine how an ensemble of denatured conformations, with a distribution
of solvent accessible areas in the DSE, gives rise to a constant $m$-value.
We show that the m-values are nearly constant for two
reasons: (1) As previously shown \cite{AutonPNAS2005,BolenPNAS2007}, the bulk of the contribution to
$\Delta G_{ND}([C])$ changes come from the protein backbone. (2) Here, we establish that  the distribution of the backbone solvent accessible
surface area is narrow, with small changes in  $\Delta \alpha_{k}^B$  as [C] decreases.

Determination of the molecular origin of denatured state collapse, often associated with
a concentration dependent $m$-value, requires characterizing the DSE of protein L at low
[C] ($<$ 3 M urea) where the NSE is thermodynamically favored. Under these
conditions we find that the radius of gyration ($R_g$) DSE undergoes significant reduction 
as [C] decreases. Urea-induced collapse transition of protein L is continuous
as a function of [C], and results in native-like secondary
structural elements. We decompose the non-bonded energy into residue-solvent
and intrapeptide interactions and show that (1) these two opposing energies govern the behavior
of $R_g$ of the DSE, and (2) the strength of these interactions
are non-uniformly distributed in the DSE and correlate with regions of
residual structure. Thus, different regions of the DSE can collapse to varying degrees
as [C] changes.
\\
\\
\\
\noindent
\textbf{Methods:}

\emph{$C_{\alpha}$-side chain model for protein L:}
In order to ascertain the conditions under which Eq. \ref{m1} is a good
approximation to Eq. \ref{m2}, we use the coarse-grained
$C_{\alpha}$-side chain model ($C_{\alpha}$-SCM) \cite{thirumPNAS2000} to represent the sixty-four residue protein L.
In the $C_{\alpha}$-SCM, each residue in the polypeptide chain is
represented using two interaction sites, one that is centered on the
$\alpha$-carbon atom and another at the center-of-mass of the
side chain \cite{thirumPNAS2000}. The potential energy ($E_P$) of a given conformation
of the $C_{\alpha}$-SCM
is a sum of bond-angle ($E_A$), backbone dihedral ($E_D$), improper dihedral ($E_I$),
backbone hydrogen bonding ($E_{HB}$) and non-bonded Lennard-Jones ($E_{LJ}$) terms ($E_P = E_A + E_D + E_C + E_{HB} + E_{LJ}$).
The functional form of these terms, and derivation of the parameters used are
explained in the supporting information of reference \cite{ObrienPNAS2008}.

Sequence information is included in the $C_{\alpha}$-SCM by using non-bonded parameters that are residue dependent.
We take into account the size
of a side chain by varying the collision diameter used
in the $E_{LJ}$ term. The interaction strength between side chains $i$ and $j$,
that are in contact in the native structure, depends on the amino acid pair and
is modeled by varying the well-depth ($\epsilon_{ij}$) in $E_{LJ}$ \cite{ObrienPNAS2008}.
Thus, the $C_{\alpha}$-SCM incorporates both sequence variation and packing effects.
Numerous studies have shown that considerable insights into protein folding can be
obtained using coarse-grained models \cite{ThirumalaiCOSB1999,ShakCR2006,OnuchicJPCB2003}, thus rationalizing the choice of the $C_{\alpha}$-SCM
in this study.

\emph{Simulation details:}
Equilibrium simulations of the folding and unfolding reaction using the $C_{\alpha}$-SCM are performed using
Multiplexed-Replica Exchange (MREX) \cite{SugitaCPL1999,RheeBJ2003} in conjunction with low friction Langevin dynamics
\cite{thirumalaiFD1997} at [C]=0. We used CHARMM to carry out the Langevin dynamics \cite{Karplus1983},
while an in-house script handles the replica exchange calculation. In the MREX simulations, multiple independent trajectories are
generated at several temperatures.
In addition to the conventional replica exchange acceptance/rejection criteria for swapping conformations
between different temperatures \cite{SugitaCPL1999}, MREX also allows exchange between replicas at the same
temperature \cite{RheeBJ2003}. Replicas were run at eight temperatures: 315, 335, 350, 355, 360,
365, 380, 400 K. At each temperature four independent trajectories were
simultaneously simulated. Every 5,000 integration time-steps the system configurations were saved for analysis.
Random shuffling occurred between replicas at
the same temperature with 50\% probability. Exchanges between neighboring temperatures were
attempted using the standard replica exchange acceptance criteria \cite{SugitaCPL1999}.
A Langevin damping coefficient of $1.0$ $ps^{-1}$ was used, with a 5 fs integration time-step.
In all, 90,000 exchanges were attempted, of which the first 10,000 discarded to allow for equilibration. All
trajectories were simulated in the canonical (NVT) ensemble.

\emph{Analysis with the Molecular Transfer Model:}
We model the denaturation of protein L by urea
using the Molecular Transfer Model \cite{ObrienPNAS2008}. Previous work \cite{ObrienPNAS2008} 
has already shown that the MTM quantitatively reproduces experimentally measured single molecule FRET 
efficiencies \cite{Haran2006,eatonPNAS2007,SchulerPNAS2007} as a function of [C] (GdmCl) for protein L and the 
cold shock protein, thus validating the methodology. The MTM combines simulations
at [C]=0 with the TM \cite{TanfordJBC1963,TanfordJACS1964}, experimentally measured
transfer free energies \cite{AutonPNAS2005,BolenPNAS2007}, and a reweighting method to predict protein
properties at any urea concentration of interest \cite{Swendsen1989PRL,KumarJCC1992,SheaJCP1998,ObrienPNAS2008}.
Our previous work has shown that the MTM accurately predicts a number of molecular characteristics of proteins
as a function of denaturant or osmolyte concentration \cite{ObrienPNAS2008}.
The MTM equation, which has the form of the Weighted Histogram Analysis Method \cite{KumarJCC1992}, is
\begin{eqnarray}
\la A([C],T)\ra &=& Z([C],T)^{-1}\sum_{l=1}^{R}\sum_{t=1}^{n_l}\frac{A_{l,t}e^{ -\beta
\{E_P(l,t,[0])+\Delta G_{tr}(l,t,[C])\} }}{\sum_{n=1}^R n_n e^{f_n - \beta_n E_P(l,t,[0])}}, \label{MTM_1}
\end{eqnarray}
where $\la A([C],T)\ra$ is the average of a protein property $A$ at urea concentration [C] and temperature
$T$, and $Z([C],T)$ is the partition function. The sums in Eq. \ref{MTM_1}
are over the $R$ different replicas from the MREX simulations, that vary in terms of temperature, and $n_l$
protein conformations from the $l^{th}$ replica. The value of
$A$ from replica $l$ at time $t$ is $A_{l,t}$, and $E_P(l,t,[0])$ is the potential energy
of that conformation at [C]=0, $\beta = 1/(k_BT)$, where $k_B$ is Boltzmann's constant.
In Eq. \ref{MTM_1}, $\Delta G_{tr}(l,t,[C]))$, the reversible
work of transferring the $l,t$ protein conformation from 0 M to [C] M urea solution,
is estimated using a form of the TM, and is given by
\begin{eqnarray}
\Delta G_{tr}(l,t,[C])) &=& \sum_{k=1}^{N_{S}} \frac{ n_k\delta g_{k}^{S}([C])}{\alpha_{G-k-G}^{S}}\la\alpha_{k}^{S}(l,t,[C])\ra + \sum_{k=1}^{N_{B}} \frac{n_k\delta g_{k}^{B}([C])}{\alpha_{G-k-G}^{B}}\la\alpha_{k}^{B}(l,t,[C])\ra. \label{ttm_3}
\end{eqnarray}
All terms in Eq. \ref{ttm_3} are the same as in Eq. \ref{m2} except instead of computing a
difference in surface areas, only the surface areas from conformation $l,t$ ($\la\alpha_{k}^{P}(l,t,[C])\ra$)
are included. In the denominator of Eq. \ref{MTM_1}, the sum is over the different
replicas and $n_n$, $\beta_n$ and $f_n$ are, respectively, the number of conformations
from replica $n$, $\beta_m = 1/(k_BT_m)$ where $T_m$ is the
temperature of $m^{th}$ replica, and the free energy $f_m$ of
replica $m$ is obtained by solving a self-consistent equation (see reference \cite{Swendsen1989PRL}).

In computing $\la\alpha_{k}^{P}(l,t,[C])\ra$ for use in Eq. \ref{ttm_3} we
use the radii listed in Table \ref{sc_size} where the backbone group corresponds
to the glycine. These parameters are different from the ones reported
in \cite{ObrienPNAS2008}. They result in better
agreement between predicted $m$-values using the MTM and predicted $m$-values from
Auton and Bolen's implementation of the TM \cite{BolenME2007,BolenPNAS2007}.
The values for $\alpha_{G-k-G}^{S}$, used in Eq. \ref{ttm_3}, are reported in Table \ref{sasa_tripep}.

We calculate the average of a number of properties of protein L using Eq. \ref{MTM_1}.
The end-to-end distance ($R_{ee}$) of a given conformation
is the distance between the $C_\alpha$ sites at residues one and sixty-four. The radius of gyration, $R_g$, is computed
using $\sqrt{R_g^2} = \frac{1}{2N-N_G}\la \sum_{i=1}^{2N-N_G}(r_i - r_{CM})^2\ra$,
where $N$ is the number of residues, $N_G$ is the number of
glycines in the sequence, $r_i$ is the position of interaction site $i$, and $r_{CM}=1/(2N-N_G)\sum_{i=1}^{2N-N_G}r_i$
is the mean position of the $2N-N_G$ interaction sites of the protein.
The solvent accessible surface area of a backbone or a side chain
($\alpha_{k}^{P}$) in residue $k$ in a given conformation
was computed using the CHARMM program \cite{Karplus1983}, which computes the
analytic solution for the surface area. A probe radius of 1.4 \AA,
equivalent to the size of a water molecule, was used.

\begin{figure}[ht]
\includegraphics[width=5.4in]{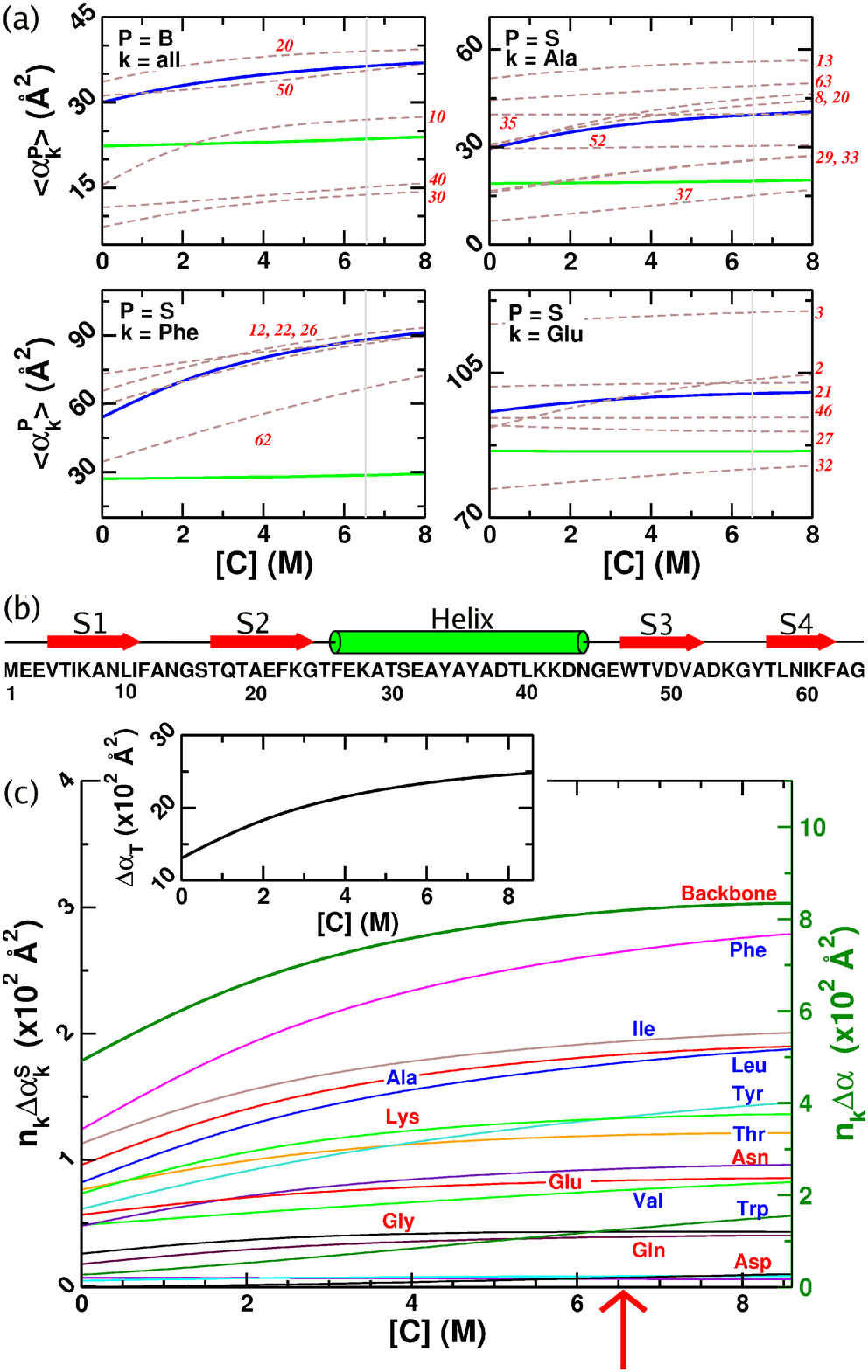}
\caption{}
\label{avg_sasa_res}
\end{figure}

The extent to which a structural element is formed (denoted $f_S$) in a conformation
of protein L is defined by $Q_p$, the
fraction of native backbone contacts formed by structural element $p$, where
$p=$ $\beta$-hairpin $S12$ or $S34$, or $\beta$-strand pairing between $S1$ and $S4$.
We define $Q_p$ as
\begin{eqnarray}
Q_p &=& \sum_j^{N-4}\sum_{k=j+4}^{N}\frac{\Theta(R_C - d_{jk})}{C_{p}}, \label{qi}
\end{eqnarray}
where the sum is over the $N=64$ $C_\alpha$ sites,
$R_C(=8$ \AA$)$ is a cutoff distance, and $d_{jk}$
is the distance between interaction sites $j$ and $k$, and $\Theta (R_C - d_{jk})$ is the
Heaviside step function.
Strand 1 ($S1$) corresponds to residues 4-11, $S2$ between 17-24,
$S3$ corresponds to 47-52, and $S4$ between 57-62 (Fig. \ref{avg_sasa_res}b).
In Eq. \ref{qi}, $C_p$ is the maximum number of native contacts
for structural element $p$. The extent of helix formation in a conformation $r$ of protein L
is computed as the ratio $N_\phi(r)/N_\phi(N)$, where $N_\phi(r)$ is
the number of neighboring dihedral pairs, between residues 26 and 44, that have
dihedral angles within $\pm 20^\circ$ of the dihedral's value in the native state, and
$N_\phi(N) = 15$.

The non-bonded interaction energy $E_I$ in the $C_\alpha$-SCM
is $E_I = E_{LJ} + E_{HB}$. We include
only the Lennard-Jones (LJ) and hydrogen bond (HB)
energies in $E_I$ \cite{ObrienPNAS2008}. The urea
solvation energy, $E_S$, of a given conformation is set equal to Eq. \ref{ttm_3}; $E_M$
is a simple sum of $E_I$ and $E_S$. The values of $E_I$ and $E_S$ for the various
structural elements of protein L were computed by neglecting non-bonded
and solvation energies of residues that were not part of the structural element
of interest.

The time-series of the various properties were inserted
into Eq. \ref{MTM_1} to compute their averages as a function of [C].
To compute averages $\la A_D \ra$ and $\la A_N \ra$ of the DSE and NSE respectively, a modification to Eq. \ref{MTM_1}
was made. The numerator was multiplied by $\Theta_n(l,t)$,
where $\Theta_n(l,t)$ is the Heaviside step function that is equal to $\Theta(5 - \Delta(l,t))$ when the
average of the NSE is computed (i.e. $n=$NSE) and is equal to
$\Theta(5 + \Delta(l,t))$ when the average of the DSE is computed
(i.e. $n=$DSE). Here, $\Delta(l,t)$ is the root mean squared
deviation between the $C_\alpha$ carbon sites in the $C_\alpha$-SCM of
conformation $l,t$ and the $C_\alpha$ carbon atoms in the crystal structure (PDB ID 1HZ6 \cite{ZhangACS2001}).
When $\Delta(l,t)$ is greater than 5 \AA\ then $\Theta(5 + \Delta(l,t))=0$ and $\Theta(5 - \Delta(l,t))=1$,
and when $\Delta(l,t)$  is less than 5 \AA\ then $\Theta(5 + \Delta(l,t))=1$ and $\Theta(5 - \Delta(l,t))=0$.

Probability distributions were computed using $P(A\pm\delta_A;[C]) = Z(A\pm\delta_A,[C],T)/Z([C],T)$, where
 $Z(A\pm\delta_A,[C],T)$ is the restricted
partition function as a function of $A$. Due to the discrete nature of the simulation data, a bin with
finite width $\pm\delta_A$, whose value depends on $A$, is used. $Z(A\pm\delta_A,[C],T) = \sum_{l=1}^{R}\sum_{t=1}^{n_l}\frac{f_A(l,t)e^{ -\beta
\{E_P(l,t,[0])+\Delta G_{tr}(l,t,[C])\} }}{\sum_{n=1}^R n_n e^{f_n - \beta_n E_P(l,t,[0])}}$, where all terms
are the same as in Eq. \ref{MTM_1} except for $f_A(l,t)$, which is a function that we define to equal 1 when the
protein conformation $l,t$ has a value of $A$ in the range of $A\pm\delta_A$, and zero otherwise.
\\
\\
\noindent
\textbf{Results and Discussion} 
\\
\\
\noindent
\textbf{$\Delta G_{ND}([C])$ changes linearly as urea concentration increases:}
We chose the experimentally well characterized B1 IgG binding domain of protein L \cite{bakerJMB2000,Haran2006,eatonPNAS2007} to illustrate
the general principles that explain the linear dependence of $\Delta G_{ND}([C])$
on [C] for proteins that fold in an apparent two-state manner. In our earlier study \cite{ObrienPNAS2008}, we
showed that the MTM accurately reproduces several experimental measurements including [C]-dependent
energy transfer as a function
of guanidinium chloride (GdmCl) concentration. Prompted by the success of the MTM, we now explore
urea-induced unfolding of protein L. The MTM predictions for urea effects are expected to
be more accurate than for GdmCl, since the experimentally measured $\delta g_{k}^{P}([C])$ urea data,
used in Eq. \ref{m1}, includes activity
coefficient corrections while the GdmCl data does not \cite{BolenPNAS2007,TanfordJBC1963}.
The calculated $\Delta G_{ND}([C])$ as a 
function of urea concentration for protein L shows linear dependence above [C] $> 4$ M
(Fig. \ref{dGns}b) with $m= 0.80$ kcal mol$^{-1}$ M$^{-1}$, and a $C_m$ (obtained using
$\Delta G_{ND}([C_m])=0$) $\approx$ 6.6 M. The consequences of the deviation from linearity, which  is observed for [C] $< 3$ M, are explored below. It should be stressed that the error in the estimated $\Delta G_{ND}([0])$ is relatively
small ($\sim$0.8 kcal mol$^{-1}$) if measurements at [C] $> 4$ M are extrapolated to [C]$=0$ (Fig. \ref{dGns}b). Thus,
from the perspective of free energy changes the assumption that $\Delta G_{ND}([C]) = \Delta G_{ND}([0])
+ m[C]$, with constant $m$, is justified for this protein.
\\
\\
\noindent
\textbf{Molecular origin of constant $m$-values}
\\
\\
\noindent
Inspection of Eq. \ref{m2} suggests that there are three possibilities that can
explain the constancy of $m$-values, thus making Eq. \ref{m1} a good approximation to Eq. \ref{m2}:
(1) Both $\la\alpha_{k,D}^{P}([C])\ra$ and $\la\alpha_{k,N}^{P}([C])\ra$ in Eq. \ref{m2}
have the same dependence on [C], making $\Delta \alpha_k^P$ effectively independent of [C].
(2) The distributions $P(\alpha_{k,D}^P;[C])$ in Eq. \ref{m2}
are sharply peaked about their mean or most probable values of $\alpha_{k,D}^P([C])$ at all
[C], thus making $\Delta \alpha_{k}^P$ independent of [C].
In particular, if the standard deviation in $\alpha_{k,D}^P$ (denoted $\sigma_{\alpha_k}$) is much less
than $\la\alpha_{k,D}^P([C])\ra$ for all [C]'s then the $\Delta \alpha_k^P$'s would be effectively
independent of [C]. (3) One group in the protein, denoted $l$ (backbone in proteins),
makes the dominant contribution to the $m$-value.
In this case, only the changes in $\Delta \alpha_{l}^P$ and $P(\alpha_{l,D}^P;[C])$
matter, thereby making $\Delta \alpha_{l}^P$ insensitive to [C].
The MTM simulations of protein L allow us to test the validity of these plausible explanations 
for the constancy of $m$-values, especially when [C]$> 3$ M (Fig. \ref{dGns}b). Only
by examining these possibilities, which requires changes in the distribution of
various properties as [C] changes, can the observed constancy of $m$ be rationalized.
\\
\\
\noindent
\textbf{$\la\alpha_{k,D}^{P}([C])\ra$ and $\la\alpha_{k,N}^{P}([C])\ra$
do not have the same dependence on [C]:}
The changes in $\la\alpha_{k,D}^P\ra$ and $\la\alpha_{k,N}^P\ra$ 
as a function of [C] show that as [C] increases, both $\la\alpha_{k,D}^P\ra$ and $\la\alpha_{k,N}^P\ra$ increase 
(blue and green lines in Fig. \ref{avg_sasa_res}a).
However, $\la\alpha_{k,D}^P([C])\ra$ has a stronger
dependence on [C] than $\la\alpha_{k,N}^P([C])\ra$ for 
both the backbone and side chains (Fig. \ref{avg_sasa_res}a). Thus, the observed linear dependence
of $\Delta G_{ND}([C])$ on [C] cannot be rationalized in terms of
similarity in the variation of $\la\alpha_{k,D}^{P}([C])\ra$ and $\la\alpha_{k,N}^{P}([C])\ra$
as [C] changes. 
The stronger dependence of $\la\alpha_{k,D}^P([C])\ra$ on [C] arises from the greater range and magnitude of the 
solvent accessible surface areas available to the DSE (see below). The greater range allows larger shifts 
in $\la\alpha_{k,D}^P([C])\ra$ than $\la\alpha_{k,N}^P([C])\ra$ with [C]. Equally important, the strength of the 
favorable protein-solvent interactions is positively correlated with the magnitude of the surface area 
and [C] (see Eq. \ref{ttm_3} and Fig. \ref{dGns}a). Thus, the DSE conformations with larger surface area are stabilized 
to a greater extent than the NSE conformations with increasing [C] and subsequently $\la\alpha_{k,D}^P([C])\ra$ 
shows a stronger dependence on [C].
\\
\\
\noindent
\textbf{Surface area distributions are broad in the DSE:} The 
variation of $\Delta \alpha_{k,D}^S$ and $\Delta \alpha_{k,D}^B$ 
with [C] suggests that the $P(\alpha_{k,D}^{P};[C])$ 
are not likely to be narrowly peaked, and must also
depend on [C] (Eq. \ref{m2}). 
As urea concentration increases, the total backbone 
surface area distribution in the DSE, $P(\alpha_D^B;[C])$,
shifts towards higher values of $\alpha^B_D$ and becomes narrower (Fig. \ref{pasa_tot_back}a).
A similar behavior is observed
in the distribution of the total surface area (Fig. \ref{pasa_tot_back}b) and for
the side chain groups (data not shown).
It should be noted that the change in $\alpha^B_D$ with [C] is about five
times smaller than the corresponding change in $\alpha_T$ (compare Figs. \ref{pasa_tot_back}a and \ref{pasa_tot_back}b).
Thus, the distribution of surface areas for the various protein components are moderately dependent
on [C], and $\Delta \alpha_T$ is more strongly dependent on [C] (Fig. \ref{avg_sasa_res}c inset).
These findings would suggest that $m$ should be a function of [C] above 4 M (Eq. \ref{m2}), in
contradiction to the finding in Fig. \ref{dGns}b.

We characterize the width of the denatured state $P(\alpha_{k,D}^P)$ distributions
by computing the ratio $\rho_k=\sigma_{\alpha_{k,D}}/\la\alpha_{k,D}^P\ra$,
where $\sigma_{\alpha_{k,D}} = \sqrt{\la {\alpha_{k,D}^P}^2\ra - \la\alpha_{k,D}^P\ra^2}$.
Fig. \ref{rho_mc}a shows $\rho_k$ as a function of [C] for the various protein components
(backbone, side chains, and the entire protein). As with the backbone $P(\alpha^B_D)$
distribution (Fig. \ref{pasa_tot_back}a), $\rho_k$ indicates that
$P(\alpha_{k,D}^P)$ becomes narrower at higher urea concentrations for most $k$ (Fig. \ref{rho_mc}a).
At 8 M urea, the width of $P(\alpha_{k,D}^P)$ ranges from 5 to 25 \% of 
the average value of $\alpha_{k,D}^P$ for all groups, except $k=Trp$ which has
an even larger width.
Clearly, $\rho_k$ is large at all [C], which accounts for the dependence of
$\Delta\alpha_k^P$ on [C]. The results in Fig. \ref{rho_mc} show that there are
discernible changes in $\rho_k$ which reflects the variations in $P(\alpha_{k,D}^P;[C])$ as 
[C] is changed.
Consequently, the constancy of the $m$-value cannot be explained 
by narrow surface area distributions.
\\
\\
\noindent
\textbf{The weak dependence of changes in accessible surface area of the 
protein backbone on [C] controls the linear behavior of $\Delta G_{ND}([C])$:} 
Plots of $m[C]$, at several urea concentrations for the entire protein, the backbone groups 
(second term in Eqs. \ref{m1} and \ref{m2}), and the hydrophobic side chains 
Phe, Leu, Ile, and Ala are shown in Fig. \ref{rho_mc}b.
The slope of these plots is the $m$-value, which in the transition region (\emph{i.e.} from 5.1 M to 7.9 M urea) 
is 0.80 kcal mol$^{-1}$ M$^{-1}$ for the
entire protein. The contribution from the backbone alone is 0.76 kcal mol$^{-1}$ M$^{-1}$,
and from the most prominent hydrophobic side chains (Phe, Leu, Ile, and Ala) is a combined 0.04 kcal mol$^{-1}$ M$^{-1}$.
Thus, the largest contribution to the change in the native state stability, as
[C] is varied, comes from the burial or exposure of the protein backbone (95\%).
The simulations directly support the previous finding that 
the  protein backbone contributes the most to the stability changes with [C] \cite{AutonPNAS2005}.
Thus, for [C] $>$ 3 M the magnitude of
the $m$-value is largely determined by the backbone groups.  However,
only by evaluating the [C]-dependent changes in the distribution of
surface areas can one assess the extent to which Eq. \ref{m2} be approximated
by Eq. \ref{m1}.

The relative change in accessible surface area of the backbone $\Delta \alpha_D^B$
has a relatively weak urea dependence between 4 M to 8 M urea, increasing by only $75$ \AA$^2$
(Fig. \ref{avg_sasa_res}c). 
Such a small change in $\Delta \alpha_D^B$ with [C] has a negligible effect on the $m$-value.
These results show that $m$ is effectively independent of [C] in the transition region because 
$\Delta \alpha_k^B([C])$ associated with the backbone groups
change by only a small amount as [C] changes, despite the fact that
$\Delta \alpha_T$ can change appreciably ($\Delta \alpha_T(4 M \rightarrow 8 M) \approx 300$ \AA$^2$ Fig. \ref{avg_sasa_res}c inset). 
Thus, the third possibility is correct, namely that the weak dependence of $\Delta \alpha_k^B([C])$ on [C]
results in $m$ being constant.
\\
\\
\noindent
\textbf{Residual denatured state structure leads to the inequivalence of amino acids:}
In applying Eq. \ref{m1} to predict $m$-values, it is assumed that
all residues of type $k$, regardless of their sequence context, 
have the same solvent accessible surface area in the DSE \cite{AutonPNAS2005,BolenPNAS2007}.
Our simulations show that this assumption is incorrect. Comparison
of $\alpha_{k,D}^P$ for individual residues of type $k$, and the average $\la\alpha_{k,D}^P\ra$
as a function of urea concentration (Fig. \ref{avg_sasa_res}a) shows that both
sequence context and the distribution of conformations in the DSE determine the
behavior of a specific residue. 
Large differences between $\alpha_{k,D}^P$ values are observed between residues
of the same type, including alanine, phenylalanine and
glutamate groups, even at high urea concentrations (Fig. \ref{avg_sasa_res}a). 
The inequivalence of a specific residue in the DSE is similar to NMR chemical shifts
that are determined by the local environment. As a result of variations in the
local environment not all alanines in a protein are equivalent.
Thus, ignoring the unique surface area behavior
of individual residues in the DSE could lead to errors in the predicted $m$-value. Because the backbone
dominates the transfer free energy of the protein (Fig. \ref{rho_mc}b), errors arising from this assumption may be small.
However, the dispersions in the backbone $\alpha_{k,D}^B$ suggests that different regions of the protein
may collapse in the DSE at different urea concentrations, driven by differences in $\Delta \alpha_{k,D}^B$
from residue to residue (see below).

The simulations can be used to calculate [C]-dependent changes in surface areas
of the individual backbone groups as well as side chains. Interestingly, even for the chemically
homogeneous backbone group, significant dispersion about $\la \alpha_{k,D}^B\ra$ is observed
when individual residues are considered (Fig. \ref{avg_sasa_res}a).
For example, $\alpha_{k,D}^B$ for residue 10 changes more drastically as [C] decreases
than it does for residues 20 or 50. Thus, the connectivity of the backbone group can not only alter the conformations as [C] is varied but also the 
contribution to the free energy. 

Even more surprisingly, the changes
in $\alpha_{k=Ala,D}^S$ depends on the sequence location of a given alanine residue
and the associated secondary structure adopted in the native conformation.
The changes in $\alpha_{k=Ala,D}^S$ for residues 8 and 20, both of which adopt  
a $\beta$-strand conformation in the native structure (Fig. \ref{avg_sasa_res}b), exhibit
similar changes upon a decrease in [C] (Fig. \ref{avg_sasa_res}a). By comparison,
surface area changes in alanine residues 29 and 33, that are helical 
in the native state (Fig. \ref{avg_sasa_res}b), are similar as [C] varies, while
the changes in $\alpha_{k=Ala,D}^S$ for alanines
that are in the loops (residues 13 and 63) are relatively small. 
Examining the probability distribution of surface areas for the individual alanines ($P(\alpha_{Ala,D}^S)$
in Fig. \ref{P_sasa}), which is related to the average surface area and higher order moments, a 
wide variability between different residues is observed.
Similar conclusions can 
be drawn by analyzing the results for the larger hydrophobic residue Phe and
the charged Glu (Fig. \ref{avg_sasa_res}a). Thus, for a given amino acid type,
both sequence context as well as the heterogeneous nature
of structures in the DSE lead to a dispersion about the average $\la \alpha_{k,D}^S\ra$
and higher order moments of $P(\alpha_{k,D}^S)$
as urea concentration changes. Much like the chemical shifts in NMR, the distribution functions of chemically identical individual residues bear signatures of
their environment and the local structures they adopt as [C] is varied!

The total surface area difference between $N$ and $D$ ($\Delta \alpha_T$)
changes by about 1,200 \AA$^2$ as [C] decreases from 8 M to 0 M
(see inset of Fig. \ref{avg_sasa_res}c). Decomposition
of $\Delta \alpha_T$ into contributions from backbone and side chains (Eqs. \ref{m1}
and \ref{m2}) shows that the burial of the backbone groups contributes the most (up to 38\%)
to $\Delta \alpha_T$ (Fig. \ref{avg_sasa_res}c). Not unexpectedly, 
hydrophobic residues (Phe, Ile, Ala, Leu), which are buried
in the native structure, also contribute significantly to $\Delta \alpha_T$, which supports the recent all atom molecular dynamics simulations \cite{BernePNAS2008}.
Among them, Phe, a bulky hydrophobic residue, makes the largest side chain
contribution to $\Delta \alpha_T$ (Fig. \ref{avg_sasa_res}c).
For example, as urea concentration increases from 4 M to 8 M the total backbone
$\Delta \alpha_{D}^B$ increases by 75 \AA$^2$, and $n_k \Delta \alpha_{k,D}^S$ for
$k=${$Phe, Leu, Ala, Ile$} increase by 21-42 \AA$^2$.

\begin{figure}[ht]
\includegraphics[width=7.0in]{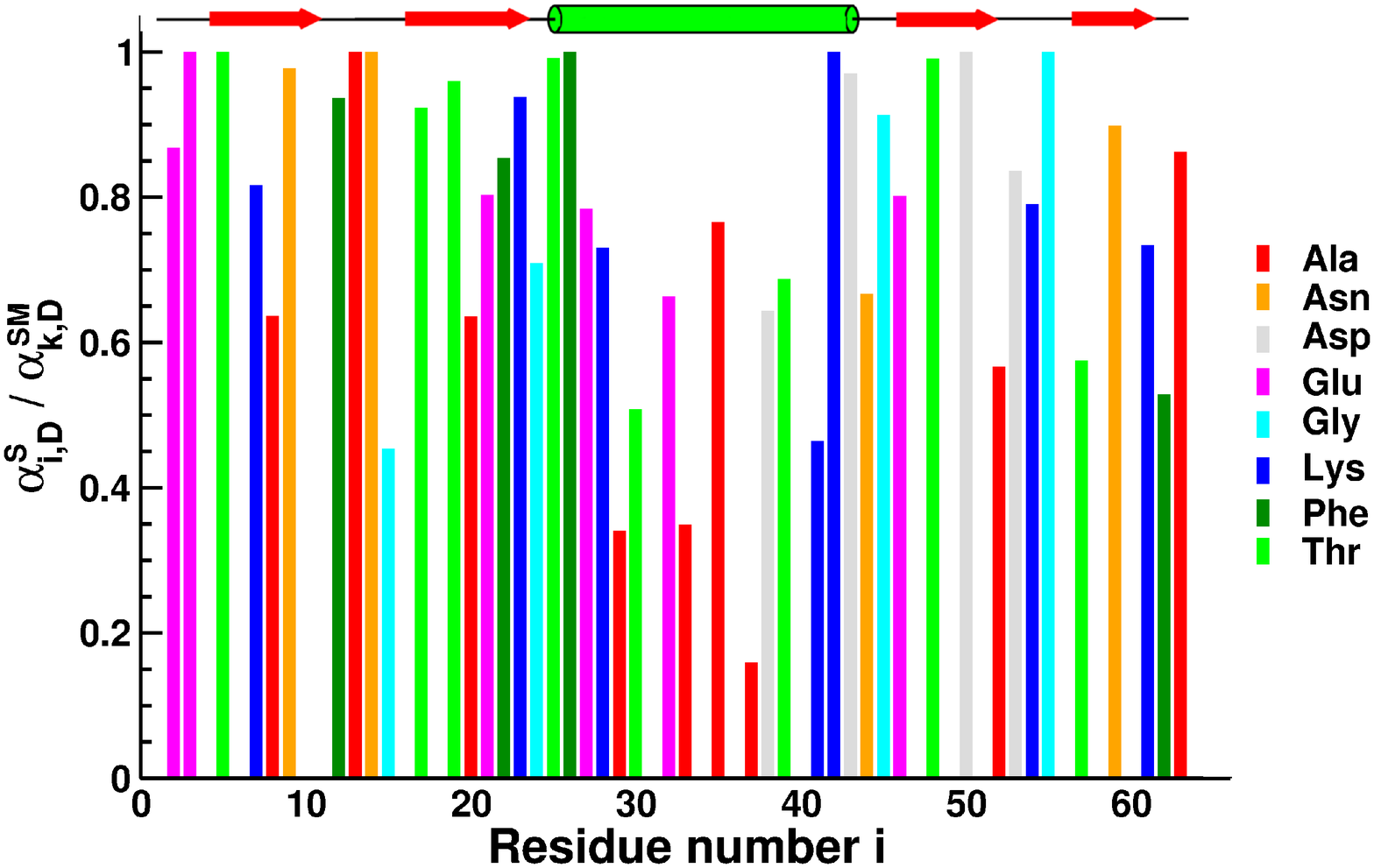}
\caption{}
\label{sasa_vs_ss}
\end{figure}

The dispersion in $\alpha_{k,D}^P$ could be caused by residual structure in the DSE \cite{BolenBIOCHEM1998,BolenPSFG2000}. We test this proposal quantitatively
 by plotting $\alpha_{k,D}^S / \alpha_{k,D}^{SM}$ for each residue, where $\alpha_{k,D}^{SM}$
is the maximum $\alpha_{k,D}^S$ value for residue type $k$ in 8 M urea. If residual structure causes
the dispersion in $\alpha_{k,D}^P$ then we expect that $\alpha_{k,D}^S / \alpha_{k,D}^{SM}$ 
should depend on the secondary structure element that residue $k$ adopts in the native state.
We find that there is a correlation between $\alpha_{k,D}^S$ and the helical secondary
structure element (residues 26 to 44, Fig. \ref{sasa_vs_ss}). The helical region tends to have
smaller $\alpha_{k,D}^S / \alpha_{k,D}^{SM}$ values compared to other regions of the protein.
Of the nine alanines  in protein L, four 
are found in the helical region of the protein. These four residues 
have some of the smallest $\alpha_{k,D}^S/\alpha_{k,D}^{SM}$ values out of the nine alanines. 
The [C]-dependent fraction of residual secondary structure in the DSE shows that
at 8 M urea the helical content is 32\% of its value in the native state (Fig. \ref{resid_inte}a).
Taken together, these data show that $\alpha_{k,D}^P$ depends not only on the residue type, 
but \emph{also} on the residual structure present in the DSE, which at all values of [C], is determined by the polymeric nature of proteins.
\\
\\
\noindent
\textbf{Residue-dependent variations in the transition midpoint - The Holtzer Effect:}
Globally, the denaturant-induced unfolding of protein L may be described
using the two state model (Fig. \ref{dGns}b). However, 
deviations from an all-or-none transition can be discerned if the residue-dependent
transitions $C_{m,i}$ can be measured. 
For strict two-state behavior, $C_{m,i}=C_m$
for all $i$, where $C_{m,i}$ is the urea concentration below which the $i^{th}$ residue
adopts its native conformation. The inequivalence of the amino acids, described above (Fig \ref{avg_sasa_res}a),
should lead to a dispersion in $C_{m,i}$. The values of $C_{m,i}$ are determined by
specific interactions, while the dispersion in $C_{m,i}$ is a finite-size effect \cite{ThirumalaiJCC2002,ThirumalaiPRL2004}.
In other words, because the number of amino acids ($N$) in a protein is finite, all
thermodynamic transitions are rounded instead of being infinitely sharp. Finite-size
effects on phase transitions have been systematically studied in spin systems \cite{FisherPRL1972}
but have received much less attention in biopolymer folding \cite{ThirumalaiPRL2004}.
Klimov and Thirumalai \cite{ThirumalaiJCC2002} showed that the dispersion in the residue-dependent melting
temperatures $T_{m,i}$,
denoted $\Delta T$ ($\Delta C$), for temperature (denaturant) induced unfolding scales
as $\Delta T/T_m \sim 1/N$ ($\Delta C/C_m \sim 1/N$). 
The expected dispersion in $C_{m,i}$ or $T_{m,i}$ is the Holtzer effect.

In the context of proteins, Holtzer and coworkers \cite{HoltzerBJ1997} were the first to observe that
although globally thermal folding of the 33-residue GCN4-lzK peptides can be described
using the two state model, there is dispersion in the melting temperature
throughout the protein's structure. In accord with expectations based on the finite size 
of GCN4-lzK, it was found, using one-dimensional NMR experiments, that $T_{m,i}$ 
depends on the sequence position. The deviation of $T_{m,i}$ 
from the global melting temperature is as large as 20\% \cite{HoltzerBJ1997}. 
More recently, large
deviations in $T_{m,i}$ from $T_m$  have been observed for other proteins \cite{MunozNAT2006}.

We have determined, for protein L, the values of $C_{m,i}$ using $Q_i(C_{m,i}) = 0.5$,
where $Q_i$ is the fraction of native contacts for the $i^{th}$ residue. The distribution of $C_{m,i}$
show the expected dispersion (Fig. \ref{holtzer}a), which implies different residues
can order at different values of [C]. The precise $C_{m,i}$ values are
dependent on the extent of residual structure adopted by the $i^{th}$ residue, 
which will clearly depend on the protein. Similarly, the distribution of the melting
temperature of individual residues $T_{m,i}$, calculated using
$Q_i(T_{m,i}) = 0.5$, also show variations from $T_m$. However, the width of the thermal
dispersion is narrower then obtained from denaturant-induced unfolding (Fig. \ref{holtzer}b). 
This result is in accord with the general observation that thermal melting is
more cooperative than denaturant-induced unfolding \cite{Thirum1998FD}.
It should be emphasized that the Holtzer
effect is fairly general, and only as $N$ increases will $\Delta C$ and $\Delta T$ 
decrease.
\\
\\
\noindent
\textbf{Specific protein collapse at low [C], and the balance between
solvation and intraprotein interaction energies:} As [C] is decreased below 3 M
there is a deviation in linearity of $\Delta G_{ND}([C])$ (Fig. \ref{dGns}b) 
and the $m$-value depends on [C]. At low [C] values the 
characteristics of the denatured state change significantly relative to the denatured state at 8 M. 
The radius of gyration $R_g^D$ and $\Delta \alpha_{T}$
change by up to 6 \AA\ (Fig. \ref{rg}) and 1,150 \AA$^2$ (Fig. \ref{avg_sasa_res}c) respectively,
indicating that the denatured state undergoes a collapse transition.
We detail the consequences of the [C]-dependent changes and examine the nature and origin of
the collapse transition.

\emph{Surface area changes:} Above 4 M urea, the $\alpha_{k,D}$ values change only modestly (Fig. \ref{avg_sasa_res}a).
However, below 4 M much larger changes in $\alpha_{k,D}$ occur (Fig. \ref{avg_sasa_res}a).
In particular, $\Delta \alpha_T$ decreases by 850 \AA$^2$ going from [C]=4 M to [C]=0 M urea, 
compared to $\approx$300 \AA$^2$ upon decreasing [C] from 8 M to 4 M urea (Fig. \ref{avg_sasa_res}c inset). The
backbone is the single greatest contributor to $\Delta \alpha_T$, accounting
for 24\% to 38\% of  $\Delta \alpha_T$ at various [C]. Thus, a significant amount of backbone
surface area in the DSE is buried from solvent as [C] is decreased, and the protein
becomes compact (Fig. \ref{avg_sasa_res}c).
The next largest contribution to $\Delta \alpha_T$, as measured by $n_k \Delta \alpha_k(= n_k(\la\alpha_{k,D}([C])\ra - \la\alpha_{k,N}([C])\ra))$,
arises from the hydrophobic residues Phe, Ile, and Ala (Fig. \ref{avg_sasa_res}c). 
These residues also exhibit relatively large changes in the DSE surface area
as [C] is decreased. 
The large change in surface area of Phe as [C] decreases shows that dispersion interactions
also contribute to the energetics of folding \cite{BernePNAS2008}.
On the other hand, for side chains
that are solvent exposed in the native state, such as the charged residue Asp, 
$n_k \Delta \alpha_k$ is small and does not change significantly with [C] (Fig. \ref{avg_sasa_res}c).
The results in Fig. \ref{avg_sasa_res}, and the surface area dependence of
the TM, suggests that the changes in
surface area at low [C] are related to changes in solvation energy of the backbone
(see below).

\emph{$R_g$ and $R_{ee}$ changes:} Decreasing [C] below 4 M leads to a $R_g^D$ change of up to 4 \AA,
and an end-to-end distance ($R_{ee}$) change of up to 10 \AA\ (Fig. \ref{rg}). 
Such a large change in $R_g^D$ shows that a collapse transition occurs in the DSE. 
We find no evidence (e.g. a sigmoidal transition
in $R_g^D$ versus [C]) that the DSE at 0 M ($\la R_g^D\ra = 15.5$ \AA) and the DSE
at 8 M urea ($\la R_g^D\ra = 21.5$ \AA) are distinct thermodynamic states. This suggests that
the urea-induced DSE undergoes a continuous second order collapse transition as urea
concentration decreases.

\emph{Residual structure changes:} To gain insight into secondary structure
changes that occur during the collapse transition
we plot the residual secondary structure ($f_S^D$) in the
DSE versus [C] (Fig. \ref{resid_inte}a). Above 4 M urea only $\beta$-hairpin 3-4 and the
helix are formed to any appreciable extent. However, below 4 M $\beta$-hairpin 1-2 and $\beta$-sheet interactions
between strands 1 and 4 can be found in the DSE. For example, at 1 M urea $\beta$-hairpin 1-2 and strands
1 and 4 are formed 21\% and 16\% of the time, while there is 56\% helical
and 74\% $\beta$-hairpin 3-4 content in the DSE (Fig. \ref{resid_inte}a). Thus, as [C] is decreased, the 
residual structure in the DSE increases, contributing to changes in $R_g$, $R_{ee}$,
 and the surface areas. This finding suggests that the collapse transition 
is specific in nature, leading to compact structures with native-like secondary structure elements.

\emph{Solvation versus intraprotein interactions:} Neglecting changes in protein conformational entropy,
two opposing energies control the [C]-dependent behavior of $R_g^D$; the interaction of
the peptide residues with solvent (the solvation energy, denoted $E_S$), and the intraprotein non-bonded 
interactions between the residues (denoted $E_I$).
For denaturants, such as urea, $E_S$ favors an increase in $R_g^D$ and a concomitant 
increase in solvent accessible surface area, while $E_I$ typically 
is attractive and hence favors a decrease in $R_g^D$. Because $E_S$ in the TM model is proportional
to a surface area term, and $E_I$ is likely to be approximately proportional to the number of residues
in contact (which increases as the residue density increases upon collapse), we expect
$E_S([C]) \propto -[C]\la R_g^D([C])\ra^2$ and $E_I([C]) \propto -1/\la R_g^D([C])\ra^{3}$. The behavior of these two functions
(increasing $\la R_g^D([C])\ra$ leads to a more favorable $E_S([C])$ and unfavorable $E_I([C])$)
suggests that there should always be some contraction (expansion) of the DSE with decreasing (increasing)
[C].
The molecular details in the $C_{\alpha}$-SCM allow us to exactly determine
$E_S([C])$ and $E_I([C])$ as a function of [C], and thereby get an understanding of the
energy scales involved in the specific collapse of the DSE.

In the inset of Fig. \ref{resid_inte}b we plot
$E_S([C])$, $E_I([C])$, and $E_M([C])$($\equiv E_S([C])$ + $E_I([C])$) in the DSE.
As indicated by the Flory-like argument given above, $E_S([C])$ becomes more
favorable with increasing [C], and $E_I([C])$ becomes more unfavorable with increasing [C]
(Fig. \ref{resid_inte}b Inset). The behavior of $E_M([C])$ is 
important to examine, as this quantity governs the behavior of $R_g^D([C])$. Above 4 M, $E_M([C])$ 
is relatively constant, varying by no more than 1 kcal/mol. This finding is consistent
with the small changes in $R_g^D$, $R_{ee}$, and $\Delta \alpha_{k}^P$ above 4 M urea (Figs. 
\ref{rg} and \ref{avg_sasa_res}c).
Below 4 M, the $E_M([C])$ strength increases and is dominated by the
attractive intrapeptide interactions ($E_I([C])$) at low [C] (Fig. \ref{resid_inte}b Inset),
driving the collapse of the protein as measured by $R_g^D$.

We dissect the monomer interaction energies further by computing the average
monomer interaction energy per secondary structural element (Fig. \ref{resid_inte}b). 
Above 4 M urea, the monomer interaction energies change by less than 0.4 $k_BT$,
except for the $\beta$-hairpin 3-4 which changes by as much as $\sim 0.9$ $k_BT$.
Below 4 M the monomer interaction energies change by as much as 1.5 $k_BT$,
with the helix exhibiting the smallest change with [C]. These
findings, which are in accord with changes in residual secondary structure (Fig. \ref{resid_inte}b),
indicate that the magnitude of the driving forces 
for specific collapse
(defined as $\frac{dE_M([C])}{d[C]}$) are (from greatest to least) associated with 
$\beta$-hairpin 3-4 $>$ $\beta$-strands 1-4 $>$ $\beta$-hairpin 1-2 $>$ helix.
Thus, the forces driving collapse are non-uniformly distributed throughout 
the native state topology.
\\
\\
\noindent
\textbf{Concluding remarks} 
\\
\\
\noindent
The major findings in this paper reconcile 
the two-state interpretation of denaturant $m$-values with the 
broad ensemble of conformations in the unfolded state, and resolves
an apparent conundrum between protein collapse and the linear variation of
$\Delta G_{ND}([C])$ with [C]. 
The success of the TM model in estimating  $m$-values \cite{AutonPNAS2005,BolenPNAS2007} suggests that the free energy of the protein can be
decomposed into a sum of independent transfer energies of backbone and side
chain groups (Eq. \ref{m1}).
However, in order to connect the measured $m$-values to the heterogeneity in the molecular conformations it is
necessary to examine how the distribution of the DSE changes as [C] changes. This requires
an examination of the validity of 
the second, more tenuous assumption in the TM, according to which the denatured ensemble surface area exposures of the backbone
and side chains  do not change as [C] changes.
This assumption, whose validity has not been 
examined until the present work, implies that neither the polymeric nature of proteins,
the presence of residual structure in the DSE, nor the extent of protein collapse alters $\la\alpha_{k,D}^P([C])\ra$
or $\la\alpha_{k,N}^P([C])\ra$ significantly. Our work shows that as urea
concentration (or more generally any denaturant) changes there are substantial changes in $P(\alpha_T)$
(Fig. \ref{pasa_tot_back}b), $R_g$, and $R_{ee}$ (Fig. \ref{rg}). 
However, because backbone groups, whose $\alpha_{k,D}^B$ values are more narrowly 
distributed than almost all other groups (see Fig. \ref{rho_mc}a), make the dominant contribution
to the $m$-value (see Fig. \ref{rho_mc}b), the $m$-value is constant in the transition region.
Therefore, approximating Eq. \ref{m2} using Eq. \ref{m1} causes only small errors in the 
range of 3 M to 8 M urea for protein L.

The utility of the TM in yielding accurate values of $m$ using measured transfer free energies of
isolated groups, without taking the polymer nature of proteins into
account, has been established in a series of papers \cite{AutonPNAS2005,BolenPNAS2007}. The success of the empirical TM (Eq. \ref{m1}), 
with its obvious limitations, has been rationalized \cite{AutonPNAS2005,BolenPNAS2007} by noting that the backbone makes the dominant
contribution to $m$. The present work expands further on this
perspective by explicitly showing that the total backbone surface
changes ($\Delta \alpha_B$) area changes weakly with [C] (for [C] $>$ 3M
for protein L).   We conclude that Eq. \ref{m1}, with the assumption that
changes in surface areas are approximately [C]-independent, is
reasonable.  This finding, to our knowledge, has not been demonstrated
previously. We ought to emphasize that $m$, a single parameter, is
only a global descriptor of the properties of a protein at $[C] \ne
0$.  Full characterization of the DSE requires calculation of changes
in the distribution functions of a number of quantities (see Figs. \ref{pasa_tot_back}a and \ref{pasa_tot_back}b) as a function of [C]. 
This can only be accomplished using MTM-like simulations and/or NMR experiments, which are by no means routine. 
The paucity of NMR studies that have characterized [C]-dependent changes
in the DSE, at the residue level, shows the difficulty in performing
such experiments.

The MTM simulations show discernible deviations from linear behavior at [C] $< 3$ M (Fig. \ref{dGns}b),
which can be traced to changes in the backbone surface area in the DSE. The structural
characteristics of the unfolded state under such native conditions are different
from those at [C] $>>$ [C$_m$]. The values of $\Delta \alpha_{k,D}^P$ are relatively 
flat when [C] $>$ [C$_m$] (Fig. \ref{avg_sasa_res}b) but decrease below [$C_m$] because of
protein collapse. 
Because $\delta g_k^B([C])$ dominates even below [$C_m$] (Fig. \ref{rho_mc}b) 
it follows that departure from linearity in $\Delta G_{ND}([C])$ is largely due to burial
of the protein backbone. The often-observed drift in baselines of spectroscopic probes
of protein folding may well be indicative of the changes in $\Delta \alpha_{k,D}^B$,
and reflect the changing distribution of unfolded states \cite{BolenBIOCHEM1988,BarrickPS2003}. 
Single molecule experiments \cite{HaasBIOCHEM2001,UdgaonkarJMB2005,Nienhaus2005PNAS,Haran2006,SchulerPNAS2007},
that directly probe changes in the DSE even below [$C_m$], exhibit large shifts
in the distribution of FRET efficiencies with [C]. Our simulations are consistent with
these observations.
The logical interpretation is that the DSE and, in particular, the distribution of
$\alpha_T$, $\alpha_B$, and the radius of gyration $R_g$ must be [C]-dependent. The present
simulations suggest that only by carefully probing these distributions, can the replacement
of Eq. \ref{m2} by Eq. \ref{m1} be quantitatively justified. In particular, large changes in the
DSE occur under native conditions. Therefore, it is important to characterize the
DSE under native conditions to monitor the collapse of proteins.

Equilibrium SAXS experiments on protein L at various guanidinium chloride
concentrations found that $R_g$ does not change significantly above [$C_m$] \cite{Baker1999Nature}. 
The $\sim$2 \AA\ change in $R_g^D$ above [$C_m$] observed in these simulations 
is within the $\approx\pm$1.8 \AA\ error bars of the experimentally measured $R_g$ above [$C_m$] \cite{Baker1999Nature}.  Our findings also suggest that 
the largest change in $R_g^D$ occurs well below [$C_m$] (3 M urea or less).
Under these conditions the fraction of unfolded molecules is less than 1\% (Fig. \ref{dGns}b inset), which
implies it is difficult to  accurately measure the $R_g$ of the DSE using current SAXS experiments and explains why the equilibrium collapse transitions
are not readily observed in scattering experiments. The present work and increasing 
evidence from single molecule FRET experiments show that the
denatured state can undergo a continuous collapse transition that is 
modulated by changing solution conditions. This finding
underscores the importance of quantitatively characterizing 
the DSE in order to describe the folding reaction.  In order to establish if the collapse transition is second order, which is most likely the case, will require tests similar to that proposed by Pappu and coworkers \cite{VitalisJMB08}.
\\
\\
\noindent
\textbf{Acknowledgments:} We thank Govardhan Reddy for a critical reading
of this manuscript. We thank Prof. Buzz Baldwin for his interest, comments and for a
tutorial on the historical aspects of the transfer model.

\newpage
\noindent
\textbf{\large{References}}
\mciteErrorOnUnknownfalse
\ifx\mcitethebibliography\mciteundefinedmacro
\PackageError{biochem.bst}{mciteplus.sty has not been loaded}
{This bibstyle requires the use of the mciteplus package.}\fi

\newpage
\begin{sidewaystable}
\centering
\caption{van der Waals radius of the side chain beads for various amino-acids based in part on measured partial molar volumes \cite{ZamyatninARBB1984}.}
\vspace{1cm}
\renewcommand{\thefootnote}{\thempfootnote}
\begin{tabularx}{35mm}{cc}
\hline
Residue & Radius (\AA) \\
\hline
  Ala & 2.14 \\
  Cys & 2.33 \\
  Asp & 2.37 \\
  Glu & 2.52 \\
  Phe & 2.70 \\
  Gly & 2.70 \\
  Hsd\footnote{The same value of the radius was used regardless of the protonation state.} & 2.63\\
  Ile & 2.63 \\
  Lys & 2.70 \\
  Leu & 2.63 \\
  Met & 2.63 \\
  Asn & 2.33 \\
  Pro & 2.36 \\
  Gln & 2.56 \\
  Arg & 2.79 \\
  Ser & 2.20 \\
  Thr & 2.39 \\
  Val & 2.49 \\
  Trp & 2.88 \\
  Tyr & 2.75 \\
\hline
\end{tabularx}
\label{sc_size}
\end{sidewaystable}
\clearpage

\begin{table}
\centering
\caption{Solvent accessibility of the backbone and side chain groups of residue $k$
in the tripeptide $Gly-k-Gly$ ($\alpha_{Gly-k-Gly}$)}
\vspace{1cm}
\renewcommand{\thefootnote}{\thempfootnote}
\begin{tabularx}{50mm}{ccccc}
\hline
\multicolumn{5}{c}{$\alpha_{Gly-k-Gly}$ (\AA$^2$)}\\
\hline
$k$ && Backbone && Side chain \\
\hline
Ala && 62.5 && 108.3 \\
Met && 50.3 && 164.7 \\
Arg && 46.2 && 186.0 \\
Gln && 52.1 && 155.4 \\
Asn && 55.6 && 138.7 \\
Gly && 85.0 && 0.0 \\
Tyr && 47.3 && 179.9 \\
Asp && 56.7 && 133.7 \\
Trp && 43.8 && 198.7 \\
Phe && 48.3 && 174.6 \\
Cys && 57.7 && 128.6 \\
Pro && 56.9 && 132.7 \\
Lys && 48.3 && 174.6 \\
Hsd\footnote{Hsd - Neutral histidine, proton on ND1 atom. Hse - Neutral histidine, proton on NE2 atom. HSP - Protonated histidine.} && 51.4 && 159.2 \\
Hse && 51.6 && 159.2 \\
Hsp && 51.4 && 159.2 \\
Ser && 60.9 && 114.9 \\
Thr && 56.2 && 135.7 \\
Val && 53.8 && 147.1 \\
Ile && 50.3 && 164.7 \\
Glu && 53.0 && 150.8 \\
Leu && 50.3 && 164.6 \\
\hline
\end{tabularx}
\label{sasa_tripep}
\end{table}
\clearpage

\newpage
\noindent
\textbf{\large{Figure Captions}}
\\
\\
{{\bf Figure} \ref{dGns}:} (a) The transfer free energy of the backbone (the
glycine residue) and side chain groups as a function of urea concentration. The
lines are a linear extrapolation of the experimentally measured $\delta g_k$ upon
transfer from 0 M to 1 M urea \cite{BolenPNAS2007}. The amino acid corresponding
to a given line is labeled using a three letter abbreviation. Blue labels
are for hydrophobic side chains, while red labels indicate polar or charged side chains
according to the hydrophobicity scale in \cite{RoseSCI1985}.
(b) The native state stability (black circles) 
of protein L as a function of urea concentration, [C], at 328 K. $\Delta G_{ND}([C])=-k_BT ln(P_N([C])/(1-P_N([C])))$,
where $P_N([C])$ is the probability of being folded as a function of [C]. 
The midpoint of the transition $C_m = 6.56$ M urea.
The red line is 
a linear fit to the data in the range of 5.1 to 7.9 M. At [C] $<$ 3 M there is a 
departure from linearity (i.e. a [C]-dependent $m$-value). Inset in the upper left
is a ribbon diagram of the crystal structure of protein L \cite{ZhangACS2001}. Inset
in bottom right shows $P_N([C])$ versus [C] at 328 K (blue line). In addition, $|dP_N/d[C]|$, the
absolute value of the derivative of $P_N$ versus [C] is shown (green line). The full width at 
half the maximum value of $|dP_N/d[C]|$ (denoted 2$\delta C$) is 2.8 M and is defined as the `transition region'
given by $C_m \pm \delta C$.

{{\bf Figure} \ref{avg_sasa_res}:} (a) $\la\alpha_{k,j}^P\ra$ versus urea concentration for
the backbone and the side chains alanine, phenylalanine, and glutamate, computed using
$\la\alpha_{k,j}^{P}([C])\ra = \int_{0}^{\infty}\alpha_{k,j}^{S}P(\alpha_{k,j}^{P};[C])d\alpha_{k,S}$ 
($j =$ $D$ or $N$ and $P =$ $S$ or $B$). For the backbone
$\la \alpha_{k=all,j}^{B}([C])\ra = N^{-1}\sum_{k=1}^{N_B}\la \alpha_{k,j}^{B}([C])\ra$,
where $N = 64$, the number of residues in the protein.
$\la\alpha_{k,N}^P\ra$ and $\la\alpha_{k,D}^P\ra$ are displayed as green and blue lines respectively.
Brown dashed lines show $\la\alpha_{k,D}^P\ra$ for individual residues of type $k$, the
residue indices are indicated by the numbers in red. For the
backbone only six groups (from residues 1, 10, 20, 30, 40, and 50) out of sixty-four 
backbone groups are shown.
(b) Linear secondary structure representation of protein L.
$\beta$-strands
are shown as red arrows, the $\alpha$-helix as a green cylinder, and 
unstructured regions as a solid black line. 
Secondary
structure assignments were made using the STRIDE program \cite{STRIDE1995}.
The residues corresponding
to each secondary structure element are listed below the representation.
(c) $n_k\Delta \alpha_k^P$ (Eq.\ref{m2}) as a function of urea
concentration for the backbone (green line, with corresponding ordinate on right), and all other sixteen unique
amino acid types in protein L (with corresponding ordinate on left). For clarity, labels for Met and Ser residues
are not shown. Met and Ser have $n_k\Delta \alpha_k^P$ values close to zero
in this graph. $\Delta \alpha_k^{P} = \la\alpha_{k,D}^{P}\ra-\la\alpha_{k,N}^{P}\ra$ ($P= S$ or $B$).
For the backbone we plot $\sum_{k=1}^{N_B}n_k\Delta \alpha_k^{B}$.
The inset shows $\Delta \alpha_T$ as a function of urea
concentration. The red arrow indicates $C_m$.

{{\bf Figure} \ref{pasa_tot_back}:} (a) The probability distribution of the total backbone surface
area in the DSE ($P(\alpha_D^{B})$) at various urea concentrations, indicated by the number
above each trace. For comparison, $P(\alpha_N^{B})$ for the native state ensemble at 6.5 M urea is
shown (solid brown line) as well as the average distribution over both the NSE and DSE
at 6.5 M urea (black line). (b) Same as (a) except distributions are of the accessible surface area
of the entire protein.

{{\bf Figure} \ref{rho_mc}:} (a) The ratio $\rho_k = \sigma_{\alpha_{k,D}}/\la\alpha_{k,D}^P\ra$
(see text for explanation)
as a function of urea concentration for the entire protein (black line), the backbone (blue line), 
and all other amino acid types found in protein L. (b) The quantity m[C] versus urea
concentration for the full protein (black circles), the backbone groups (red squares),
and the Phe, Leu, Ile, and Ala side chains.
Solid lines correspond to linear fits to the data in the range of 5.1 to 7.9 M urea.

{{\bf Figure} \ref{P_sasa}:} The distribution ($P(\alpha_{Ala,D}^S)$) of the solvent accesible surface area of side chains from the nine individual
alanine residues in the denatured state ensemble of protein L at various urea concentrations.
Black, red and green lines correspond to 1 M, 4 M and 8 M urea respectively. The corresponding
alanine for each graph is given by its residue number. The large changes in ($P(\alpha_{Ala,D}^S)$) for the chemically identical residue shows
that environment and local structures affect the structures and energetics of the side chains.

{{\bf Figure} \ref{sasa_vs_ss}:} The ratio $\alpha_{i,D}^S / \alpha_{k,D}^{SM}$ (see text for
an explanation) as a function of
residue number $i$ at 8 M urea. The legend indicates the amino acid type for each residue.
Only amino acid types that occur at least four times in protein L, and 
have at least two of those residues separated by more than twenty five residues along
sequence space, are plotted.  For reference,
the linear secondary structure representation of protein L is shown
above the graph.

{{\bf Figure} \ref{resid_inte}:} (a) The residual secondary structure content in
the DSE versus urea concentration. (b) The interaction energy ($E_M$)
in the DSE divided by the number of residues in the secondary structural element, in units of $k_BT$, 
versus urea concentration for the entire protein and
various secondary structural elements. The inset shows $E_I$, $E_S$, and $E_M$ for the entire protein
versus urea concentration in units of kcal mol$^{-1}$.

{{\bf Figure} \ref{holtzer}:} The histogram of residue-dependent midpoints of unfolding as a
function of (a) urea concentration at 328 K and (b) temperature at 0 M urea. The
$C_m$ for the entire protein is $\sim$6.6 M, while the melting temperature is
356 K at 0 M urea.

{{\bf Figure} \ref{rg}:} The average $R_g$ (open black circles)
and $R_{ee}$ (x's) as a function of [C] for protein L at 328 K. The values of $R_g^{DSE}$ (open black circles, dashed line, left axis)
and $R_{ee}^{DSE}$ (x's, dashed line,
right axis) as a function of urea concentration are also shown. Lines are
a guide to the eye. The gray vertical line at 6.56 M urea denotes the $C_m$.

\begin{figure}[ht]
\includegraphics[width=6.5in]{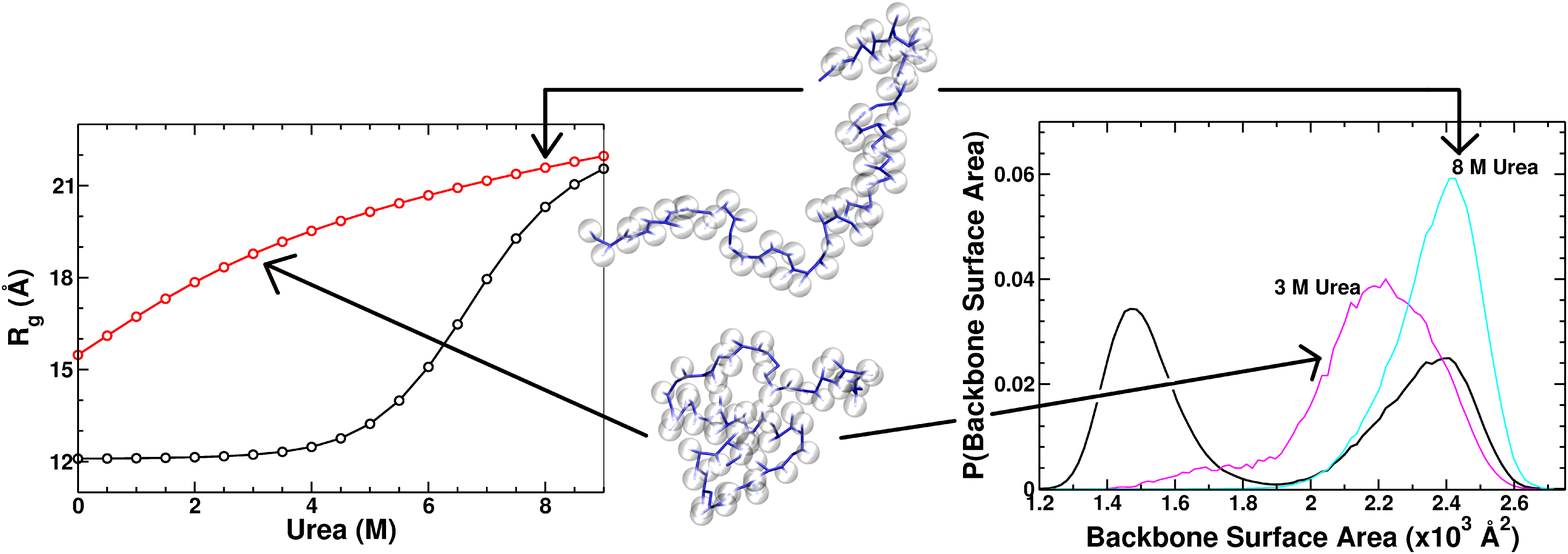}
\caption{}
\label{toc}
\end{figure}

{{\bf Figure} \ref{toc}:} Table of contents graphic.

\newpage

\clearpage

\begin{figure}[ht]
\subfigure[]{
  \label{}
  \includegraphics[width=4.7in]{fig3a.eps}
} \\
\vspace{1cm}
\subfigure[]{
  \label{}
  \includegraphics[width=4.7in]{fig3b.eps}
}\\
\caption{}
\label{pasa_tot_back}
\end{figure}
\clearpage

\begin{figure}[ht]
\subfigure[]{
  \label{}
  \includegraphics[width=4.7in]{fig4a.eps}
} \\
\vspace{1cm}
\subfigure[]{
  \label{}
  \includegraphics[width=4.7in]{fig4b.eps}
}\\
\caption{}
\label{rho_mc}
\end{figure}
\clearpage

\begin{figure}[ht]
\includegraphics[width=5.4in]{fig5.eps}
\caption{}
\label{P_sasa}
\end{figure}
\clearpage

\clearpage

\begin{figure}[ht]
\subfigure[]{
  \label{}
  \includegraphics[width=4.7in]{fig7a.eps}
} \\
\vspace{1cm}
\subfigure[]{
  \label{}
  \includegraphics[width=4.7in]{fig7b.eps}
}\\
\caption{}
\label{resid_inte}
\end{figure}
\clearpage

\begin{figure}[ht]
\subfigure[]{
  \label{}
  \includegraphics[width=4.7in]{fig8a.eps}
} \\
\vspace{1cm}
\subfigure[]{
  \label{}
  \includegraphics[width=4.7in]{fig8b.eps}
}\\
\caption{}
\label{holtzer}
\end{figure}
\clearpage

\begin{figure}[ht]
\includegraphics[width=5.4in]{fig9.eps}
\caption{}
\label{rg}
\end{figure}
\clearpage

\clearpage

\end{document}